%% file: __MAIN__.tex
\documentclass[10pt,lettersize,journal]{IEEEtran}
\usepackage{amsmath,amsfonts}
\usepackage{algorithmic}
\usepackage{algorithm}
\usepackage{graphicx}
\usepackage{textcomp}
\usepackage{xcolor}
\usepackage[caption=false,font=normalsize,labelfont=sf,textfont=sf]{subfig}
\usepackage{textcomp}
\usepackage{stfloats}
\usepackage{verbatim}
\usepackage{cite}
\usepackage{tabularx,booktabs}
\usepackage{multirow}
\usepackage{multicol}
\usepackage{enumerate}
\usepackage{enumitem}
\usepackage{ragged2e}
\PassOptionsToPackage{hyphens}{url} 
\usepackage{hyperref}
\begin{document}

\title{E-MPC: Edge-assisted Model Predictive Control}

\author{
    Yuan-Yao Lou,
    Jonathan Spencer,
    Kwang Taik Kim,
    Mung Chiang
    
}



\maketitle

\begin{abstract}
    Model predictive control (MPC) has become the de facto standard action space for local planning and learning-based control in many continuous robotic control tasks, including autonomous driving. MPC solves a long-horizon cost optimization as a series of short-horizon optimizations based on a global planner-supplied reference path. The primary challenge in MPC, however, is that the computational budget for re-planning has a hard limit, which frequently inhibits exact optimization.
    Modern edge networks provide low-latency communication and heterogeneous properties that can be especially beneficial in this situation.
    We propose a novel framework for edge-assisted MPC (E-MPC) for path planning that exploits the heterogeneity of edge networks in three important ways: 1) varying computational capacity, 2) localized sensor information, and 3) localized observation histories. Theoretical analysis and extensive simulations are undertaken to demonstrate quantitatively the benefits of E-MPC in various scenarios, including maps, channel dynamics, and availability and density of edge nodes. The results confirm that E-MPC has the potential to reduce costs by a greater percentage than standard MPC does.
\end{abstract}


\input{1-Introduction}
\input{2-Problem-Statement}
\input{3-System-Framework}
\input{4-Analysis}
\input{5-Evaluation}
\input{7-Conclusion}
\input{8-Proof}

\bibliographystyle{IEEEtran}
\bibliography{reference}

\end{document}

%% file: 1-Introduction.tex
\section{Introduction} \label{sec:introduction}

Autonomous driving is the focus of a massive effort by both academia and industry to provide safe mobility in the real world. The dynamic environment, time-sensitive actions, and massive data produced by driving necessitate robust learned controllers. For many continuous robotic control tasks, especially autonomous driving, model predictive control (MPC) has become the de facto standard action space for learning-based control \cite{b1}. However, due to the real-time computational demands of MPC, the latency and bandwidth constraints of most communication systems mean that MPC must be performed entirely onboard the robot/vehicle. Nevertheless, recent improvements in communication have produced systems that enable a rethinking of classical computational structures in learning-based control. Here we discuss the potential for a new paradigm of Edge-assisted MPC (E-MPC) where an agent may benefit from the computational resources and localized information of other computational nodes.

\begin{figure}[ht!]
\centering
    \centering\includegraphics[width=0.485\textwidth]{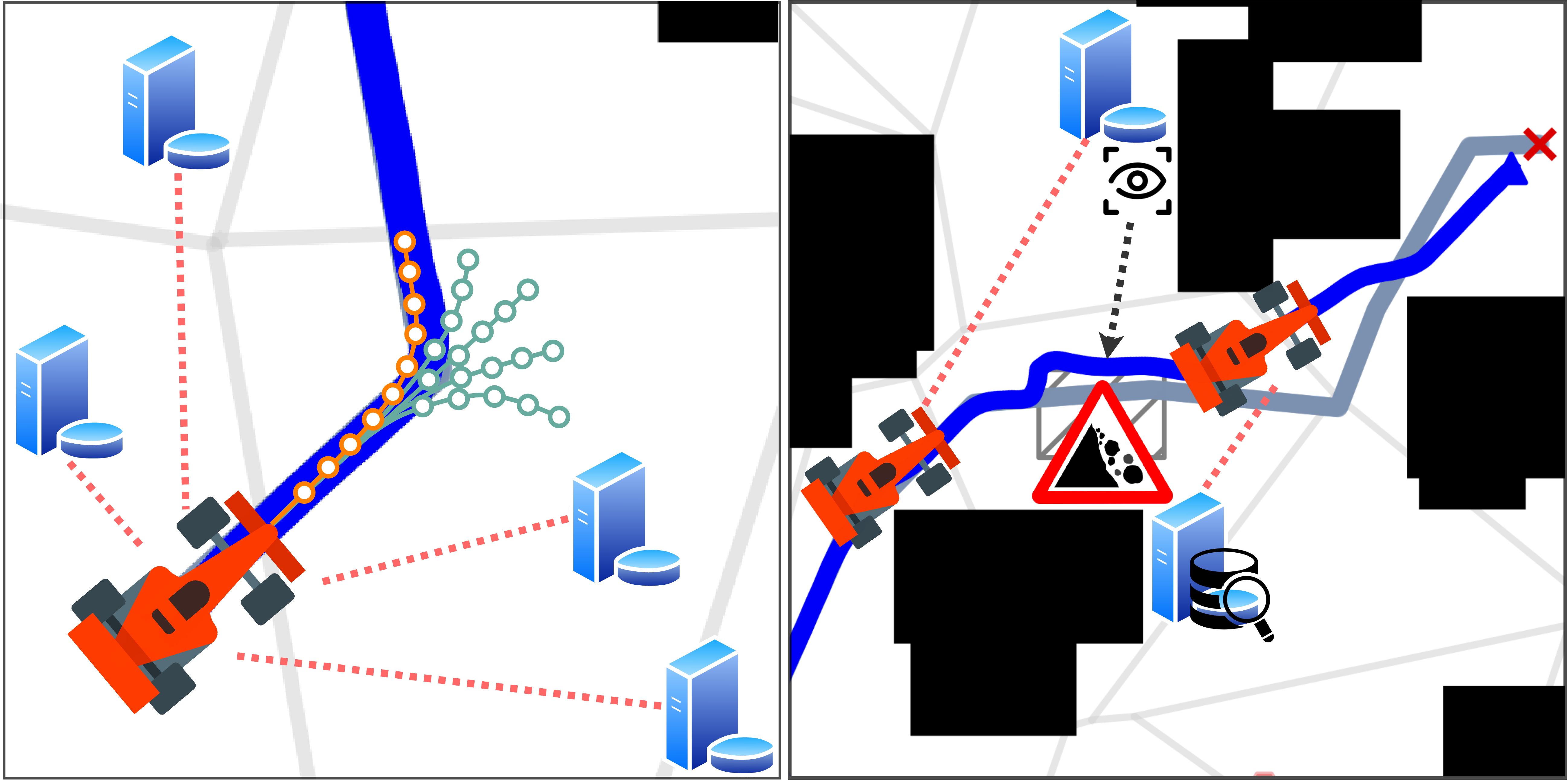}
    \caption{The examples of how edge nodes can assist MPC. Under assistance from one or multiple edge servers, the agent precisely tracks the dotted reference path, avoids unknown environment changes, and deviates from the reference path for a shortcut based on historical data.}
    \label{fig:empc-main}
\vspace{-1pt}
\end{figure}

MPC is a method of path planning that solves a long-horizon cost optimization \cite{b2} as a series of short-horizon optimizations. Whereas dynamic programming begins at the terminal state and looks backward, recursively searching for optimal control sub-sequences, MPC looks forward from the current state for the cost-minimizing control sequence within a short horizon of length $H$, executes the first action, and then re-plans. This forward optimization is based on a known (or a learned) cost function, while an approximate model of the system dynamics is used for forward prediction. The long planning horizon with frequent re-planning compensates for minor inaccuracies in the model and predictions to produce a trajectory that is close to, but not necessarily optimal.

MPC is especially effective in learning-based control. Learning a structured cost function and then evaluating short-horizon plans within that cost function leads to much stabler policies than directly regressing low-level controls. Additionally, the frequent re-planning accounts not only for model inaccuracies regarding system evolution dynamics but also for dynamic agents that may act in unpredictable ways.

\subsection{Motivations and Challenges}
Although MPC solves a shorter horizon optimization that is easier than global optimization, the large continuous state spaces and hard replanning time constraints ($\sim$10Hz) of real-time robotics mean that an exact solution is often not possible. In this case, practitioners often use motion-primitive libraries, sampling-based methods, or sparse graph search to generate a set of candidate trajectories \cite{b1} which are evaluated against a known cost. The sparse graph search method uses geometric splines of short repeated action sequences to connect the current position to a set of potential future nodes. For each future node, finding the action sequence is the problem of finding the spline parameters. In a motion-primitive library, each primitive consists of a sequence of controls $[a_1,...,a_H]$, which are fed through the planner to generate a sequence of predicted states. Because the number of possible primitives is exponential in $H$, the library is often pruned to a finite dictionary of sequences that exhibit desired characteristics such as smoothness and uniqueness. Alternatively, sampling-based methods use the entire dictionary of primitives but randomly sample a small subset of action sequences that can be evaluated quickly. Once the set of candidate primitives and corresponding predicted trajectories is defined, each trajectory is evaluated against the cost function, and the lowest-cost trajectory is selected for execution.

We are most interested in sampling-based methods for three reasons. First, sampling methods are flexible in computational cost. If more computational time and resources are available, more samples can be generated and evaluated, yielding on average better solutions. This flexibility is perfectly suited to the heterogeneity inherent to edge computing. Second, sampling methods allow for the natural incorporation of priors to focus sampling density in regions of the highest interest, improving both the efficiency and accuracy of the state estimates and thus providing better solutions. Finally, sampling methods provide a natural way to offload computation to other nodes with minimal communication and coordination overhead, compared with the other two aforementioned solutions.

In this paper, we propose a novel framework to support MPC operating in the edge computing networks for optimizing path planning, which we refer to as E-MPC, as depicted in Figure \ref{fig:empc-main}. Our method exploits the underlying variability of edge networks in three important ways.

\begin{enumerate}[leftmargin=*]
    \item   \textbf{Varying computational capacity}:
    Edge nodes of varying computational capacity can be positioned so that more
    resources are allocated to the positions of higher importance. The higher
    computational capacity of edge nodes assists agents in optimizing path
    planning. With more computing resources, sampling-based MPC can evaluate more
    candidate primitives and predict more trajectories, calibrating a more optimized path.
    
    \item   \textbf{Localized sensing information}:
    Edge nodes can use localized sensing information to improve state estimates and
    thus offer better solutions. As more localized sensing information passes into
    the sampling process, a more optimized primitive dictionary can be formed.
    Further, the results of each step in the look-ahead prediction process are also
    optimized.

    \item   \textbf{Localized observation histories}:
    Edge nodes can use localized observation histories as a prior for sampling
    motion primitives more densely in the regimes that have historically been traversed the most.
\end{enumerate} 

Our proposed approach gracefully handles degradation due to the additional latency delays and uncertainty in a wireless communication environment in two ways. First, because the agent continuously performs a local MPC computation, it can operate safely and independently without the help of edge nodes in the case of loss of communication. Second, the computation offloading of sampling-based methods only takes minimal communication overhead, thus the short transmission latency and high reliability lessen the impact of uncertainty.

\subsection{Related Work}
First introduced in the late 1970s, MPC is very well-studied and has become the de facto standard in many control settings where a reasonably good system model exists \cite{camacho2013model,rawlings2017model}. The original MPC formulation consists of a single controller and a single plant, however many variants consider the case when there are multiple of either the controller or the plant, and a cooperative (information shared between multiple controllers) or non-cooperative setting depending on the information sharing schemes between them. This distributed MPC lends itself well to tasks such as multi-vehicle formation control or collision avoidance \cite{dunbar2006distributed, christofides2013distributed}. 

We operate within a very particular setting of distributed MPC consisting of a primary controller (the agent/vehicle), the plant (the environment), and a set of auxiliary controllers (the edge nodes). This setting is highly relevant to real-time control tasks such as autonomous driving, although the benefits of edge/cloud assistance in MPC is a relatively new and unstudied. This has been studied at a high level in terms of offloading aspects of the control problem to the cloud \cite{givehchi2014control, heilig2015cloud, vick2016model}, when the cloud nodes cooperate in the presence of partial information \cite{skoko2017cloud,b4} as well as considering communication and privacy implications \cite{zhang2021privacy}.

Recent literature proposes cloud-edge-combined or edge-based MPC control systems. However, these studies either purely focus on optimizing the motion control parameters of unmanned ground vehicle (UGV) systems with a specific cost function or solely rely on remote computation resources for demonstrating the prototype system's feasibility, without considering the latency constraints imposed by wireless channel dynamics and the reliability concerns that come with it \cite{aarzen2018control, cavanini2021lpv, seisa2022edge, yang2023cloud}.

The most relevant work related to this is the recent work by Skarin et. al. who studied assisted MPC in the non-cooperative cloud setting where an agent computes locally while also querying multiple cloud nodes for computational assistance \cite{b5,b6}. In their work, the cloud nodes have identical knowledge of the objective function and state information but may vary in MPC look-ahead horizon and connection strength. They focus primarily on the effect of connection failure, choosing from available cloud solutions arbitrarily when multiple arrive in time. Our work shares a similar setting but reasons more explicitly about cost by using a probabilistic approach. Rather than varying the MPC horizon of the edge node solutions, in our approach, each edge node solves the same objective in a randomized sampling approach described in Section \ref{sec:problem_statement}. The random sampling-based approach to MPC is possibly less common than solving exactly over the horizon, however, it allows for graceful scaling in terms of edge network availability and node heterogeneity.

\subsection{Outline and Summary of Contributions}
Our contributions can be summarized as follows:
\begin{itemize}[leftmargin=*]
    \item We introduce a novel framework of assistive control called edge-assisted model predictive control (E-MPC) 
    that exploits the unique characteristics of edge networks (Sec. \ref{sec:system-framework}).
    
    \item We identify and analyze three different ways that the heterogeneity of edge networks can be utilized by
    the agent (Sec. \ref{sec:analysis}).

    \item We evaluate different server loads in a multi-agent scenario to estimate how many edge servers need to be deployed to fulfill service-level requirements. (Sec. \ref{sec:analysis-multi}).
    
    \item We demonstrate empirically that E-MPC provides significant performance advantages and show the relative
    effectiveness of each method within an autonomous driving context (Sec. \ref{sec:evaluation}).
\end{itemize}

%% file: 2-Problem-Statement.tex
\section{Problem Statement} \label{sec:problem_statement}
In this section, we first introduce the MPC local planning cost-minimization problem. (Sec. \ref{sec:cost_minimization}). 
We then provide an overview of our proposed edge-assisted MPC methodology (Sec. \ref{sec:e-mpc}).

\subsection{MPC Local Planning Cost-Minimization Problem} \label{sec:cost_minimization}
We first consider a robotic agent performing a $T$-length cost-minimization problem over sequences of states $s_{1:T}$ and actions $a_{1:T}$ for some fixed, known (though potentially learned) cost function $C(s,a)$, approximate system dynamics $\mathcal{P}$, and distribution over initial states $d_0$. The goal is to minimize the expected cost-to-go of the trajectory $\tau = s_1, a_1, ..., s_T, a_T$ from the initial state.
\begin{align} \label{eqn:goal}
    & \underset{a_{1:T} \in A}{\text{minimize}}\ C(\tau) := \underset{\substack{s_1 \sim d_0\\s_{t+1} \sim
    P(\cdot|s_t,a_t)}}{\mathbb{E}} \Bigg[ \sum_{t=1}^{T} C(s_t, a_t)\Bigg].
\end{align}
When $T$ is very large, it is easier to solve (\ref{eqn:goal}) approximately using the method of
MPC (sometimes referred to as receding horizon control), which minimizes the cost over a shorter horizon $H \ll T$, executes the first action of the short horizon minimization, then successively minimizes and executes subsequent actions based on updated state. With the MPC policy, the total cost of the trajectory is
\begin{align} \label{eqn:cost}
    & C(\tau) = \sum_{t=1}^{T} \mathbb{E}_{s_t \sim P} [C(s_t,a_t)], \nonumber \\
    & \text{where} \ a_t \in \underset{a_{t:t+H-1}}{\arg\min} \ \mathbb{E}_{s_h \sim P} \Bigg[
    \sum_{h=t}^{t+H-1} C(s_h, a_h)\Bigg].
\end{align}
Because the horizon $H$ in (\ref{eqn:cost}) is much shorter, we can simplify the computation using a fixed dictionary of $K$ motion primitives
$D_\alpha = \{\alpha^{(1)},...,\alpha^{(i)},...,\alpha^{(K)}\}$, where each primitive is a valid sequence of $H$ actions $\alpha^{(i)} = a_1^{(i)}\, ..., a_H^{(i)}$.
Alternatively, we may say that the primitives are sampled from a distribution, which may potentially depend on the state $\alpha^{(i)} \sim p(\alpha|s)$,
but in the simplest case $p(\alpha)$ is just the uniform distribution over primitives.

The cost for executing a primitive from an initial state $s_t$ is readily computed using the system model $\mathcal{P}$.
For a deterministic system, $\alpha^{(i)}$ produces a unique state sequence $s_{t:t+H}^{(i)}$, and primitive cost is $C(\alpha^{(i)}|s_t) = \sum_{h=0} ^{H-1} C(s_{h+t}^{(i)},a_h^{(i)})$
For a stochastic system, we may sample several state sequences and evaluate the average primitive cost as
$C(\alpha^{(i)}|s_t) =
            \mathbb{E}_{s^{(i)'} \sim P} \big[ \sum_{h=0} ^{H-1} C(s_{h+t}^{(i)},a_h^{(i)}) \big]$
\footnote{Depending on the application, for stochastic MDPs we may consider both mean and variance of primitive cost.}. Using the primitive library, the choice of $a_t$ from (\ref{eqn:cost}) becomes
\begin{align} \label{eqn:action}
a_t=a_0^{(i)} \in \underset{\alpha^{(i)} \in D_\alpha}{\arg\min}~ C(\alpha^{(i)}|s_t).
\end{align}

\subsection{Edge-Assisted Model Predictive Control (E-MPC)} \label{sec:e-mpc}
With this formulation, let us now consider the presence of assistive edge nodes with computing and sensing capabilities. We assume that the edge nodes are provided with both the cost function $C$ as well as the system model $\mathcal{P}$. The discrete-time index corresponds to the system’s re-planning period $T_{\nonumber \text{replan}}$, on the order of 50-100 ms. We assume that the edge nodes are communicating with the agent on a low-latency connection such that at every time step the agent can broadcast its current state estimate, and the edge nodes can perform some computation, and then feedback response to the agent within the re-planning period. When this setting\footnote{Sec. \ref{sec:analysis-conn} goes into greater depth.} holds, we propose a method for E-MPC that aids the agent in the following ways.

\subsubsection{\textbf{Increased Computational Capacity}} \label{sec:benefit-1}

The base contribution for edge-assisted MPC is that of supplemental computational capacity for computing the cost-minimizing action sequence $\alpha$. However, this should be performed in a way that does not fail catastrophically with connection failures or delays. This can be achieved when we solve \eqref{eqn:cost} using random sampling rather than exact minimization, since sampling methods can scale trivially depending on resource availability.

For our sampling approach, we assume a large fixed dictionary $D_\alpha$, $|D_\alpha|=K$ of primitives known a priori by both the agent and all edge nodes. At each time step, the agent broadcasts their current state, and all nodes begin randomly drawing from $D_\alpha$, varying in number depending on individual computational capacity and link latency. Here we treat the $n$ draws that each node performs as draws from a uniform multinomial distribution $\mathcal{M}(n,1/K,...,1/K)$. A little before the re-planning period ends, each edge node transmits the index and associated cost of their best candidate primitive. The agent collects the samples and aggregates the cost along with its local computation, choosing the minimum over the whole set. This process is repeated at every time step.

With this aggregation approach, the agent is agnostic to which of the edge nodes performed the computation. It simply observes an expanded set of cost computations from which to choose the minimum. This is great because there is no difference in performance as different nodes move in and out of range. In expectation, drawing more samples will always lead to improved performance, which diminishes with the number of additional samples as we show in Section~\ref{sec:analysis}.

\subsubsection{\textbf{Localized Sensor Data}} \label{sec:benefit-2}
The first property speaks only to the computational capacity of edge nodes and speaks nothing of their other important properties. In that regard, the agent is just as well served by a set of virtual cloud nodes running in parallel. The unique contribution of an edge network is the heterogeneity of information available at each node in the form of localized sensor data.

Consider the case where the map contains features not observable by the agent such as icy road conditions or when the map is occluded around corners. For safety reasons, we force the agent to act cautiously and assume these unobserved costly regions fill the whole map. Edge nodes dispersed across the map can provide localized sensor information that updates that information based on their local observation of conditions. As the agent streams their state, the edge node can build an improved state estimate $\hat{s_t} = \hat{s}_t^{\text{edge}} \bigcup \hat{s}_t^{\text{agent}}$ and base their cost computation estimate on that. Because the agent computes their own cost cautiously and because we assume perfect state information, the agent is guaranteed not to do worse by utilizing local information.

\subsubsection{\textbf{Localized Observation Histories}} \label{sec:benefit-3}
The assumption that the agent samples uniformly over the set of primitives, or selects from a pruned dictionary with roughly uniform coverage is perhaps naive since it may waste precious computation time performing predictions and cost estimates for primitives it is certain will have a high cost. However, in the presence of dynamic agents and a system model of degraded fidelity, it may be risky to do otherwise and rule out potentially optimal actions. This is especially the case when the agent is traversing a map for the first time and has no meaningful priors over the cost, such that
$\mathbb{E}_{\alpha \sim p(\alpha)} [C(\alpha)] = \mathbb{E}_{\alpha \sim p(\alpha|s)} [C(\alpha)]$.

However, an edge node that is positioned within the environment near a localized set of states may have observed (and computed) many cost-optimal primitive selections for a given state. Let us assume that at each time-step $t$, the agent broadcasts the previous cost-optimal primitive $\alpha_{t-1}^*$ along with the current state $s_t$ so that the edge node accumulates a local database of states and a corresponding set of primitives $D_{s,\alpha} = \{(s,\{\alpha_j\})_i\}$ which were cost-optimal within the set of primitives considered at the time. For a given state, the edge node can now build a sampling distribution around the action sequences. For a state ${s_i}$, let $h_i$ be the historical set of indices of prior min-cost samples from that node that were transmitted to the agent. 

This set $h_i$ forms a natural update for the previously uniform sampling prior. Let $p_j$ be the probability of choosing index $j$, then the probability is updated according to the number of occurrences of that index in the set $h_i$.
\begin{equation}
    p_j = \frac{(1+\sum_{a_{i'} \in h_i}  1[a_{i'}=j])}{K+|h_i|}. \nonumber
\end{equation}

As more data is collected, the prior becomes stronger and stronger, sampling most densely on previous min-cost indices. We compare the relative contribution of each of these benefits experimentally in Section~\ref{sec:evaluation}.

%% file: 3-System-Framework.tex

\section{System Framework} \label{sec:system-framework}
\begin{figure}[htbp]
    \centerline{\includegraphics[width=.35\textwidth]{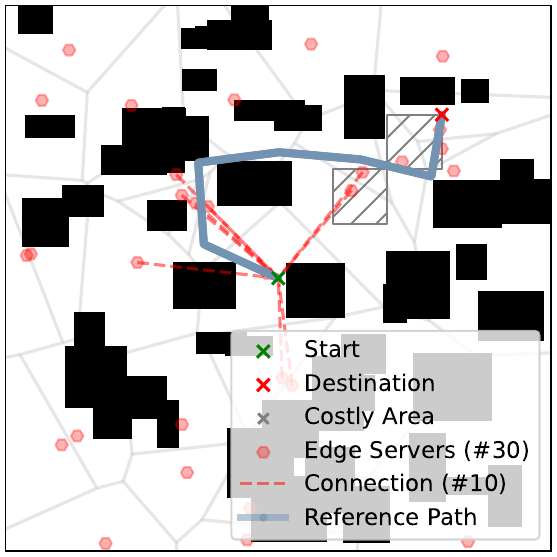}}
    \caption{The high-level overview of E-MPC system framework: the black boxes are the obstacles in the surrounding areas, while the costly areas represent the agent's uncharted environment.}
    \label{fig:system}
    \vspace{-1pt}
\end{figure}

In this section, we introduce the system framework of E-MPC in the 5G/NextG mobile network with edge computing infrastructure. We begin by elaborating on the experimental simulation of E-MPC and then explaining the overall setup and the impact of performance regarding the cost-to-go and the assistance from edge nodes. Finally, we describe the operation of E-MPC from both the client (agent) and server perspectives in light of numerous agent scenarios. Figure \ref{fig:system} depicts a high-level overview of the E-MPC system.

\subsection{Experimental Simulation} \label{sec:exp-simulation}


We develop a simulated autonomous driving framework following a client-server model and referring to the MuSHR/ROS platform for the motion control model. To elaborate on the experimental simulation, we start by introducing the generation of the motion-primitive library (Sec. \ref{sec:sys-motion}). Once the motion model and the motion-primitive library are ready and shared among the agents (clients) and the edge nodes (servers), we generate a global reference path for each unique map using the rapidly exploring random trees (RRT*) algorithm (Sec. \ref{sec:sys-global}). Then for each agent’s move following the global reference path, we evaluate the cost of all sampled trajectories (Sec. \ref{sec:sys-trajectory}). However, the setup mentioned above only suffices for the standard operation of E-MPC using onboard computing resources. To apply E-MPC to more realistic scenarios, various parameters and settings are involved and defined for different use cases in the evaluation section (Sec. \ref{sec:evaluation})


\subsubsection{\textbf{Motion-primitive library}} \label{sec:sys-motion}


In addition to the parameters of the MuSHR kinetic motion model, the MPC setting contains other key inputs that play an important role in the generation of a motion-primitive library. Specifically, the MPC parameters decide the size of look-ahead windows $H$ (i.e., the length of a control sequence, the number of actions per control sequence), the segments (i.e., the number of sets of control sequences), the branches (i.e., the number of control sequences per set), and the possible directions within a range of vision, etc. Based on the input kinetic model, these parameters are transformed into numerous trajectories, which consist number of control sequences (i.e. motion primitives $\alpha$), forming a motion-primitive library $D$. In the developed simulation, we assume the client and the server share and use the same motion-primitive library. The settings and the library generated in our simulation are defined and elaborated in Sec. 5.1, and the illustration is depicted in Fig. \ref{fig:mplib}.

\subsubsection{\textbf{Global reference path by RRT* algorithm}} \label{sec:sys-global}

Each map is unique and is created using different random seeds with 50 distributed obstacles whose size is also randomly decided. Accordingly, we provide a fixed global reference path for the agent to navigate from a starting point to a destination on each map. The reference path is established using the RRT* algorithm to sample possible paths around neighbor regions following the MPC manner to reach the final destination while avoiding obstacles. Note that although a reference path generated by RRC* is not an optimal one, it provides a feasible route in different levels of sub-optimality by different parameters (e.g., search ranges, node reconnection criteria, etc.). Given that maps are element-wise paired with global reference paths and all of the information is shared among clients and servers, the parameters are out of the scope of this work. In short, a reference path serves only as a guideline or a baseline for agents to navigate.

\subsubsection{\textbf{Trajectory sampling and cost evaluation}} \label{sec:sys-trajectory}

As mentioned in Sec. \ref{sec:problem_statement}, the motion-primitive library stores numerous trajectories according to the parameter settings. In this work, for each move using the sampling-based MPC control method, an agent randomly samples a collection of trajectories from the motion-primitive library using a uniform distribution. The number of trajectories that are sampled within a collection depends on the computation capacity. Given that a typical re-planning rate could be ten milliseconds (ms) for realistic robot control operations, the default computation capacity of an agent is set to be able to sample 10 trajectories, perform data pre-processing, and evaluate them within the re-planning rate. Note that since an edge server is typically more powerful than an end-user, as analyzed in Sec. \ref{sec:analysis-comp}, it can draw more samples to assist agents in calibrating more precise trajectories.

Each of the trajectory samples has a parameter-defined look-ahead window forming future steps as a predictive control sequence from the sampling position. These steps are processed to be mapped with a specific segment of the global reference path for cost evaluation. The rationale of an optimal trajectory is defined as following the global reference path as closely as possible to reach the destination as soon as possible while avoiding obstacles and hard steering control. Thus, to evaluate the set of candidate primitives and the corresponding predicted trajectories, the state-action cost function is a linear feature weighting $C(s, a) = w^Tf(s, a)$. The feature weights $w$ are highly sensitive and hand-tuned so that the agent approximately tracks the reference path while producing a desirable trajectory in terms of smoothness and efficiency. The features $f(s, a)$ that make up the cost function consist of several real-value evaluations, including Fréchet distance, the magnitude of steering controls, distance from the destination and obstacles, collision probability, and extra penalty. The mathematical form of $f(s, a)$ can be referred to in \cite{PROOF}.

With the above six metrics, we apply a heuristic weight vector to them in two cases which is whether the agent has a clear sight of the destination or not. Although the weights are simulation-specific hyperparameters that are hand-tuned for each map, they can be learned and adjusted by the edge servers based on the collected data from the environment and the agent. However, the methods for efficiently learning and adjusting these cost weights are outside the scope of this paper.

\subsection{E-MPC Operation} \label{sec:sys-empc}

In general, agents are provided with a reference path to the destination that begins at their starting position. A global path planner provides the reference path, which acts as a baseline for agents to evaluate the motion primitive. The agent then executes MPC local planning for each move based on a known cost function that can be improved with localized data. In E-MPC, both agents and edge servers are equipped with the aforementioned information, with the edge servers able to enhance the cost function based on data collected from themselves and connected agents.

In contrast to standard MPC, the operation of E-MPC incorporates wireless communication between agents and edge servers in order to collectively solve the MPC local planning cost-minimization problem. Each time an agent moves toward its destination, it broadcasts its current state and the previous cost-optimal primitive to all connected edge servers. After receiving this information, an edge server executes sampling-based MPC using a motion-primitive dictionary, which is fixed and shared among agents and servers. By referencing the agent’s motion model, the overall server-side execution time of E-MPC, including sampling, cost evaluation, and data transfer, is constrained by a fixed re-planning rate (i.e., 10 ms in this work). Within this time constraint, depending on the computational capacity $c$, the edge server samples multiple candidate trajectories from the dictionary and iteratively evaluates the cost of each candidate. A cost-optimal trajectory candidate is then found and sent to the agent, together with the motion primitive, state estimate, and associated costs. Finally, the agent compares the primitive proposed by the edge with the one chosen by itself to determine its next move.

In addition to considering the unpredictability of wireless communication when optimizing E-MPC, we also analyze two common scenarios that an agent can encounter. First, when surrounded by impediments, an agent may enter a blind area. This event is defined in our framework as the point at which all predicted trajectory candidates collide with impediments. Accordingly, any feasible route from the starting point to the destination with more blind spots has a significantly higher cost than others. Moreover, the environmental conditions, such as icy roads, mud areas, construction zones, etc., should also be considered from a safety standpoint while solving a forward optimization. However, agents may be unaware of these factors due to the time-varying changes. In E-MPC, we classify places with safety concerns as costly since a candidate trajectory that passes through these areas would incur higher costs.

%% file: 4-Analysis.tex
\section{Theoretical Analysis} \label{sec:analysis}

To validate the proposed benefits of the E-MPC framework, in this section, we analyze the performance of computational offloading in the E-MPC and describe how the new features of 5G NR (new radio) networks can meet E-MPC connection needs (Sec. \ref{sec:analysis-comp}, \ref{sec:analysis-conn}). Furthermore, we investigate the performance improvement brought by the localized sensing information and the cost-optimal data stored in the localized observation histories (Sec. \ref{sec:analysis-local}, \ref{sec:analysis-data}). Finally, we extend to a multi-agent scenario and use the M/M/1 queue to evaluate the deployment of edge servers against service-level requirements (Sec. \ref{sec:analysis-multi}).

\subsection{Performance Analysis of Computational Offloading} \label{sec:analysis-comp}

Here we provide an analysis of the anticipated performance improvement from the first benefit of E-MPC: computational offloading. Let us consider the case proposed where the agent and the edge nodes all share a very large dictionary of primitives of size $K$ with a uniform prior for sampling from the dictionary. Let $C_i, ..., C_K$ be random variables for the costs associated with each of those primitives. The distribution of $C_{1:K}$ will be different for every state depending on the map configuration, so here we consider two extreme cases.

In wide open areas where the primitives gradually sweep towards an obstacle, the cost map will appear as a smooth gradient and the costs will appear uniformly distributed with some constant bias and scaling: $C_i \sim (c_{max}-c_{min})*U(0, 1)+c_{min}$. In that case, if the agent samples $M$ candidate primitives, the expected cost of the best candidate is
\begin{align} \label{eqn:exp-cost1}
    \mathbb{E} \bigg[ \underset{\alpha_i \in D_\alpha}{\text{min}} \ C(\alpha_{i}|s_t) \bigg] &= \mathbb{E} [\text{min}\{C^{(1)}, ..., C^{(M)}\}] \nonumber \\ &=\frac{(c_{max}-c_{min})}{M+1}+c_{min}. 
\end{align}

Alternatively, in a case where there is a very narrow corridor blocked by obstacles, the cost map will be very low cost ($c_{min}$) for some small fraction $\gamma$ of the primitives and high cost ($c_{max}$) for the rest according to some unknown function of the primitive index $f(i)$ which without loss of generality we treat as a step function in $i$. For a single sample, the expected cost is
\begin{align} \label{eqn:exp-cost2}
    \mathbb{E} [C] = \frac{1}{K}\sum_{i=0}^{K-1} p(i) f(i) &= \frac{1}{K}\sum_{i=0}^{\gamma K-1} c_{min} + \frac{1}{K}\sum_{i=\gamma K}^K c_{max} \nonumber\\ &= (1-\gamma) c_{max} + \gamma c_{min}, 
\end{align}
where $\gamma K$ is assumed to be an integer.
For $M$ samples, using geometry the expected cost is computed to be
\begin{align} \label{eqn:exp-cost3}
    \mathbb{E} [\text{min}\{C^{(1)}, ..., C^{(M)}\}] &= (1-\gamma)^M c_{max} \nonumber \\& + (1-(1-\gamma)^M) c_{min}. \nonumber
\end{align}

In the uniform case, the distance to the optimal cost is approximately inversely proportional to the number of samples, while in the corridor case, the distance to the optimal cost is exponential in $M$ and varies depending on $\gamma$. Although this analysis is quite simplified, it shows the approximate way in which additional samples from the edge server benefit the performance. We can see that for every doubling of samples we approximately half the distance to the minimum cost. We corroborate this analysis empirically in Section~\ref{sec:evaluation}.

\subsection{Low Latency and High Reliability Requirement of E-MPC Connectivity} \label{sec:analysis-conn}

\label{sec:commrequirement}
We posit that network latency between the agent and the radio access network (RAN) can be a significant contributor to end-to-end (E2E) latency, although achievable E2E latencies depend on the available network and computing infrastructure, software features, and how the use case is implemented,
because, for time-sensitive applications such as E-MPC, 5G networks are likely to be deployed with edge nodes on-premises or at the network's edge \cite{automated-buses}. 
Consequently, we will assess the E-MPC connectivity requirements by focusing on the RAN.
Bounded low latency in RAN is addressed through numerology scaling \cite{TR38.913,Vihriala--Zaidi--Venkatasubramanian2016}. 
For instance, sub-dividing a slot further into sub-slots, rapid HARQ retransmission protocols with round-trip time scaling down in accordance, and instant transmission mechanisms to minimize the waiting time for uplink data.
Prioritization and pre-emption are also introduced by 5G as part of delivering priority and faster radio access to URLLC traffic \cite{Sachs--Andersson--Araujo2019}.

These additional features allow the base station with its scheduler to choose from a wide range of options to achieve QoS in terms of latency and reliability.
For instance, 5G NR one-way air-interface delay \cite{time-critical-comm} for downlink can achieve approximately 2ms for initial transmission (with 99\% reliability), 6ms for first retransmission (with 99.9\%), 9ms for second retransmission (with 99.99\%), and 11ms for third retransmission (with 99.999\%) for mid-band (e.g., 3.5GHz).
The latency can be lowered much further for mmWave.
The less latency there is in the connection, the more candidate primitives and predicted trajectories there are at an edge node and the better MPC local planning at an agent works. 
In the following section, we will use this approach to undertake simulations demonstrating that E-MPC has the potential to outperform standard MPC in multiple aspects.

\subsection{Analysis of Localized Sensing Information} \label{sec:analysis-local}

Although agents and servers share the same information (e.g., motion-primitive library and cost functions), as mentioned in Sec. \ref{sec:sys-empc}, the edge servers are able to improve the cost function based on the localized sensing data to reflect timely environment changing, such as icy roads, mud areas, and construction regions. Since the agents cannot recognize the latest map update, they pessimistically assume the high-cost region covers the entire surrounding environment. However, the edge servers can establish such improved state estimate $\hat{s_t} = \hat{s}_t^{\text{edge}} \bigcup \hat{s}_t^{\text{agent}}$ by combining the state received from agent $\hat{s}_t^{\text{agent}}$ and the state observed at the edge $\hat{s}_t^{\text{edge}}$. Thus, the edge servers are able to evaluate the improved state estimates $\hat{s_t}$ using the enhanced cost function $\hat{C}(s, a) = w^Tf(\hat{s}, a)$. Then by sampling from the motion-primitive library, the choice of $a_t$ from (\ref{eqn:cost}) becomes $$a_t=a_0^{(i)} \in \underset{\alpha^{(i)} \in D_\alpha}{\arg\min}~ \hat{C}(\alpha^{(i)}|\hat{s}_t).$$

The total cost of the trajectory evaluated upon improved state estimates from edge servers is guaranteed to be lower than the one assessed by the agent.
\begin{align}
    & C(\tau) = \sum_{t=1}^{T} \mathbb{E}_{s_t \sim P} [\hat{C}(s_t,a_t) \leq \sum_{t=1}^{T} \mathbb{E}_{s_t \sim P} [C(s_t,a_t)].
\end{align}

\subsection{Cost-optimal Prior Data and Trajectory Sampling} \label{sec:analysis-data}

As discussed in Sec. \ref{sec:benefit-3}, the library becomes stronger and stronger as more cost-optimal prior data is collected. This section first evaluates different map discretization methods for generating anchor points to store cost-optimal prior data (Sec. \ref{sec:data-map}). We then perform data pre-processing to address large state space issues for utilizing cost-optimal trajectory libraries (Sec. \ref{sec:data-lib}). Based on these settings, we generate cost-optimal trajectory libraries with higher sampling density in regions of the highest interest (Sec. \ref{sec:data-traj}). Finally, we apply the weight parameter $\alpha$ to the uniform and the empirical distribution and analyze the performance (Sec. \ref{sec:data-perf}).

\subsubsection{\textbf{Map discretization and anchor points}} \label{sec:data-map}

We first define an anchor point $A$ to store cost-optimal prior data in the Cartesian coordinate system. Then we define a map consisting of numerous anchor points, such that $\textit{MAP} = \{ A = (x, y)\ \lvert\ 0 \leq x \leq 1, 0 \leq y \leq 1; x \in \mathbb{R}, y \in \mathbb{R} \}$. The cost-optimal prior data is collected using the state $s_i$ of each agent's move from all successful trips over the past, forming a cost-optimal trajectory dictionary based on the prior driving results. $$D_\textit{prior} = \{ (s_i, \alpha^{(j)})_i\ \lvert\ \alpha^{(j)} = a_1^{(j)}\, a_2^{(j)}\, ..., a_H^{(j)} \}$$

Note that the size of cost-optimal trajectory dictionary $D_\textit{prior}$ depends on the number of prior trips collected in the system and the number of steps that an agent passed by on each map. Besides, based on different map discretization methods (e.g., rounding up x-y coordinates to different precision of floating point numbers), the number of anchor points to store cost-optimal prior data on a map varies. Specifically, if we choose more precise mapping, there are more anchor points distributed within an edge server’s connection range, and vice versa. On the other hand, a finer map discretization with more anchor points over the same amount of collected data reduces the size of stored data in each anchor point, decreasing the probability of a successful match.

\subsubsection{\textbf{Dimension reduction of library state space}} \label{sec:data-lib}

Given that a pose consists of three floating points (i.e., x-y coordinates and steering angles), when conducting a cost-optimal trajectory library, we only take the first two fields (i.e., x-y coordinates) to define anchor points for two reasons. First, a large state space is difficult for an exact pose mapping if the amount of collected data is not enough. Moreover, as we mentioned earlier, a large state space increases the number of anchor points, reducing the probability of a pose match. Second, in the autonomous driving context, it is natural to have automatic steering control assistance from the computation nodes (e.g., onboard computing, and edge servers). Further, we round up the x-y coordinates to the third precision of the floating point to increase the matching probability. The mapping function $f_\textit{mapping}: \mathbb{R}^3 \rightarrow \mathbb{R}^2$ is then defined as $$f_\textit{mapping}(s_i) = A_i, \forall s_i \in D_\textit{prior}, \forall A_i \in \textit{MAP}.$$

To generate a cost-optimal motion-primitive library using the prior data, we gather successful trips from various experiments and exclude the top 10\% of data with the highest and lowest cost-to-go. Then from each trip, we collect the index of the selected trajectory in each agent’s move (i.e., state), accumulating a dictionary whose key is the output of the mapping function $f_\textit{mapping}(s_i)$ and the value is a set of prior min-cost samples of motion-primitive $h$. $$D'_\textit{prior} = \{ (A, h)_i\ \lvert\ h = \{\alpha_k^{(j)}\} \}$$

Note that the size of cost-optimal trajectory dictionary $D_\textit{prior}$ also depends on the definition of the mapping function now since it transforms state information into an anchor point to gather and store cost-optimal data.

\subsubsection{\textbf{Trajectory sampling of cost-optimal library}} \label{sec:data-traj}

Different from a default motion-primitive library $D_\alpha$ where all primitives share the same probability of being sampled, a cost-optimal library $D'_\textit{prior}$ has a higher sampling density in regions of the highest interest (i.e., low-cost trajectory from prior data). That said, the more often a trajectory is selected in the prior data, it has a higher probability of being sampled in the cost-optimal library. For each anchor point with its stored prior data, we form an empirical distribution by accumulating the number of occurrences according to the stored prior data. The sampling probability of each trajectory is then calculated by dividing the number of occurrences by the total sample size. $$p(\alpha^{(j)} \lvert A_i \in D'_\textit{prior}) = \frac{\sum_{\alpha_{i'} \in h_i}  1[\alpha_{i'}=j]}{|h_i|}.$$

However, completely relying on the empirical distribution for sampling can harm the agent especially when the amount of prior data is not sufficient. Thus, we adopt a parameterized weight approach to form a distribution by the mix of the probability mass function (PMF) of the default motion-primitive library in uniform distribution $P_U$ and the PMF of the empirical distribution $P_E$. Finally, the sampling PMF from the cost-optimal motion-primitive library is defined as $$P_\textit{optimal} = P_U(u) \cdot (1 - \beta) + P_E(e) \cdot \beta,$$ where $\beta$ is a user-defined weight parameter. When $\beta$ is set to 0, the sampling follows the uniform distribution in default. In contrast, if $\beta$ is set to 1, the sampling distribution becomes the empirical distribution built from the prior data. Generally, the optimal weight parameter should be located between zero and one and keep changing according to the confidence level of the collected prior data.

\subsubsection{\textbf{Performance analysis}} \label{sec:data-perf}

Ideally, all of the anchor points should store sufficient cost-optimal prior data to ensure low cost and high reliability. However, agents should keep moving even if they arrive at an anchor point without historical information. In the E-MPC framework, an agent always performs standard MPC using onboard computing resources despite not receiving the response from the edge servers. Consequently, the motion primitives are sampled from different distributions depending on the state, such as
\begin{equation} \label{eqn:sampling}
    \underset{\alpha^{(i)} \in D_\alpha} {p(\alpha^{(i)} \lvert s)}, \\ \text{where}
    \begin{cases}
        \alpha^{(i)} \sim P_\textit{optimal}, \text{if } f_\textit{mapping}(s) \in D'_\textit{prior} \\
        \alpha^{(i)} \sim p(\alpha) = P_\textit{uniform}, \text{otherwise}.
    \end{cases}
\end{equation}

When sufficient prior driving data is collected, the total cost of the trajectory by sampling from the cost-optimal library is expected to be lower than taking actions from (\ref{eqn:action}). Specifically, $$\underset{\alpha^{(i)} \sim p(\alpha^{(i)} \lvert s_t), f_\textit{mapping}(s_t) \in D'_\textit{prior}}{C(\alpha^{(i)} \lvert s_t)} \leq \underset{\alpha^{(i)} \sim p(\alpha)}{C(\alpha^{(i)} \lvert s_t)}.$$ The size of the cost-optimal library to improve sampling effectiveness varies among anchor points and depends on their surrounding environment (refer to Appendix \ref{sec:appendix} and full proof \cite{PROOF} in more detail). Also, $P_\textit{optimal}$ is equivalent to the uniform distribution when $\beta$ is set to zero as default when the anchor point has no prior data. The minimum cost can thus be found exhaustively by sweeping $beta$ values although it requires extremely high computation resources. By sampling from the cost-optimal motion-primitive library, the choice of action control from (\ref{eqn:cost}) becomes $$a_t=a_0^{(i)} \in \underset{\alpha^{(i)} \in D_\alpha, \alpha^{(i)} \sim p(\alpha^{(i)} \lvert s_t)}{\arg\min}~ C(\alpha^{(i)}|s_t).$$

\subsection{\textbf{Server availability in multi-agent scenario}} \label{sec:analysis-multi}

In the above analysis, the edge servers are dedicated to serving one client, thus the average server load can be simply modeled as a Bernoulli distribution with a specified probability. However, in reality, the edge cloud region contains multiple users and thus the servers need to process varying numbers of requests simultaneously. We extend the prior analysis by adding one more complexity to server availability to evaluate how many edge servers need to be deployed to fulfill the service-level requirements.


We model the average server load using an M/G/1 queuing system with two parameters, $\lambda$ the arrival rate, and $\mu$ the service rate \cite{MG1queue}. Given $\lambda$ and $\mu$, the utilization rate $\rho$ is defined as $\lambda / \mu$. Then following the Pollaczek–Khinchine formula combined with Little's law \cite{PKformula}, the mean number of customers in system $L$ and the mean sojourn time $W$ are given by \begin{equation}\label{eqn:sojourn}L = \rho + \frac{\rho^2 + \lambda^2\text{Var}(S)}{2(1 - \rho)}, W = \frac{\rho + \lambda \mu \text{Var}(S)}{2(\mu - \lambda)} + \mu^{-1}.\end{equation} In the context of E-MPC, the sojourn time represents the average delay time (i.e., queuing delay plus service time), and the service rate $\mu$ of edge servers follows the exponential distribution. Therefore, the variance of service time distribution $S$ can be plugged in with $1/\mu^2$, and the mean number of customers in system $L$ and the mean sojourn time $W$ in (\ref{eqn:sojourn}) of an M/M/1 queue model become \begin{equation}L = \frac{\rho}{(1 - \rho)}, W = \frac{L}{\lambda} = \frac{1}{\mu - \lambda}.\end{equation}

To embed the average delay time into the E-MPC framework, we run an additional analytical simulation using the M/M/1 model. We first create a set of servers all of which have a series of requests and a series of service times for each request, where request arrivals occur at rate $\lambda$ according to a Poisson process and service times have an exponential distribution with rate parameter $\mu$ \cite{MM1queue}. Then we run the analytical simulation for 20 seconds to extract the first 10 seconds, which is equal to 1,000 steps in our simulation, to characterize more variances instead of capturing the converged or saturated behavior. The output from the analytical simulation is a collection of delay time (ms) at different steps in the simulation. The size of the collection is the number of deployed edge servers.

Since we also consider the re-transmission time caused by the dynamics of wireless communication as mentioned in Sec. \ref{sec:analysis-conn}, compared to Bernoulli distribution which returns a boolean output, the delay time estimated using M/M/1 analytical simulation plays a more crucial role in determining performance. Specifically, the agent discards the response from the server when the total elapsed time (i.e., delay time plus re-transmission time), resulting in different cost minimization levels. $T_{\text{delay}} + T_{\text{re-tx}} \leq 10\ \text{(ms)}.$ Last but not least, we can experimentally evaluate how many servers need to be deployed to fulfill service requirements (i.e., cost-to-go, probability of failure) under different server loads $\mathcal{L}$ conditions modeled by M/M/1 parameters $\lambda$ and $\mu$.
This result extends to a processor-sharing context, where a server’s processor tasked with handling numerous jobs tends to service these jobs on a time-sharing basis. This is due to the fact that distribution of time in a processor-sharing system becomes identical to the distribution characteristics of the M/M/1 system \cite{PKformula}.

%% file: 5-Evaluation.tex
\section{Numerical Evaluations} \label{sec:evaluation}


To demonstrate the benefits of E-MPC along with the corresponding performance gain, we design a set of simulations to test the impact of different key factors. In this section, we present an overview of the E-MPC simulation setup (Sec. \ref{sec:eval-setting}) and describe in detail the six distinct simulation scenarios (Sec. \ref{sec:eval-scenario}). Lastly, we measure performance in terms of total accumulated cost, averaged across a collection of randomly created maps, and present the corresponding numerical results with visualized and intuitive explanations (Sec. \ref{sec:eval-result}).

\begin{figure}[ht!]
\centering
    \centering\includegraphics[width=0.35\textwidth]{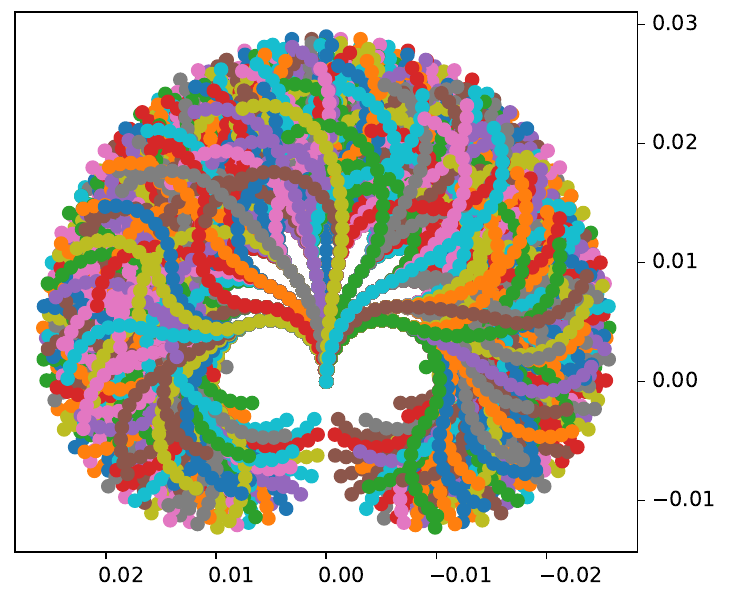}
    \caption{Motion primitive dictionary.}
    \label{fig:mplib}
\vspace{-1pt}
\end{figure}

\subsection{Setting} \label{sec:eval-setting}

The task of the agent is to navigate from a start position to a goal position as quickly as possible while avoiding collision with the randomly generated obstacles on the map. For every map, the agent is provided with an approximate reference path from a global path planner that uses the RRT* algorithm to generate a sub-optimal shortest path as described in Sec. \ref{sec:sys-global}. As mentioned in the previous section, due to the hard limit of the re-planning rate, the default computation capacity $c$ of an agent is set to 10 (i.e., 10 samples per 10 ms). The system dynamics model $P$ is a simplified kinetic model on the agent to simulate the actuator, which we adapt from MuSHR/ROS platform and integrate into our simulator \cite{srinivasa2019mushr}. The agent and the edge nodes share a motion primitive dictionary $D_{\alpha}$ of size K = 1000, depicted in Figure 3. At each state the primitive $\alpha$ is combined with motion model $P$ to produce a predicted candidate trajectory defined in Sec. \ref{sec:sys-motion}.

At each discrete time-step, the agent broadcasts the current state information to all of the connected edge servers. In the simulation, the connection range between an agent and an edge server is limited according to the distance from the server and whether if any obstacles between the agent and the server. The agents, however, always have excellent wireless channel quality if they are within the coverage of the edge servers. Then the following operation of E-MPC is described in Section \ref{sec:sys-empc}. The definition and the values of the parameters involved in different experiments are listed in Table \ref{table:parameter}.

\begin{table}[ht]
  \setlength\tabcolsep{9pt}\renewcommand\defaultaddspace{1.0ex}
  \begin{tabularx}{\columnwidth}{@{}c>{\hsize=0.8\columnwidth}X>{\hsize=0.0\hsize}X @{}}
    \toprule
    \textbf{Symbols}    & \textbf{Definition and Values} \\
    \midrule
    \addlinespace
    $c$                 & Computational capacity of edge servers \newline (Values = 0, 10, 20, 40, 80, 100) \\
    \addlinespace
    \hline
    \addlinespace
    $\delta$            & Server connection criteria as E2E latency setting \newline (Values = 0, 1, 2, 3, 4) \\
    \addlinespace
    \hline
    \addlinespace
    $N$                 & Number of deployed edge servers \newline (Values = 0, 5, 10, 15, 20, 25, 30, 35, 40, 45, 50) \\
    \addlinespace
    \hline
    \addlinespace
    $\mathcal{L}$       & Models of average server loads \newline (Values = $\textit{Bernoulli}(p)$, $\lambda$ and $\mu$ of M/M/1 queue) \\
    \addlinespace
    $p$                 & Server availability as probability in Bernoulli distribution \newline (Values = 0, 0.2, 0.4, 0.6, 0.8, 1.0) \\
    \addlinespace
    $\lambda, \mu$      & M/M/1 queue model: Arrival rate, Service rate \newline (Values = [(10, 100), (20, 200), (10, 200), (50, 300), (20, 300), (80, 400), (50, 400), (100, 500), (80, 100)]) \\
    \addlinespace
    \hline
    \addlinespace
    $\Omega$            & Number of collected cost-optimal prior data (iterations) \newline (Values = 180, 540, 900, 1260, 1800, 2340, 2700, 3060, 3600) \\
    \addlinespace
    $\beta$            & Weights for uniform and empirical distributions \newline (Values = 0.0, 0.125, 0.25, 0.375, 0.5, 0.625, 0.75, 0.875, 1.0) \\
    \addlinespace
    \hline
    \addlinespace
  \end{tabularx}
  \caption{Definitions and values of the parameters involved in the experiments.}
  \label{table:parameter}
\end{table}

\subsection{Scenarios} \label{sec:eval-scenario}

\begin{enumerate}[leftmargin=*]
    \item   \textbf{Increased computational capacity}:
    To evaluate the performance improvement in terms of different levels of assistance from the edge servers, we assign a few different fixed aggregate sample budgets for all edge servers in different runs of this scenario. The quantity of the assigned computational capacity $c$ represents the maximum number of predicted trajectory candidates on which an edge server is able to perform E-MPC within a static re-planning rate. The values of computational capacity $c$ range from 0 to 100 as defined in Table \ref{table:parameter}, in which zero value is the standard MPC. Since the agent is agnostic to the source of the samples, this sample budget can be achieved either by a single, powerful, and reliable edge node, or by a collection of edge nodes with varying computational capabilities and link quality that in aggregate produce that number of samples. As discussed in Sec. \ref{sec:analysis-comp}, the cost is expected to decrease as more computing resources are available.
    
    \item   \textbf{Wireless communication}: 
    The previous scenario only demonstrates a one-to-one client-server model with perfect conditions for wireless communication. In the following scenarios, we extend the client-server model to many-to-one so that the agent can receive assistance from multiple edge servers. When considering multiple servers in E-MPC, more factors that affect the performance are involved, such as the density and the availability of edge servers. Furthermore, the impact of performance from varying computational capacity and the dynamics of the wireless channels are considered in the following scenarios. The computational capacity $c$ of each edge server is randomly assigned in different runs of the simulations with values from 5, 10, 20, to 40.
    \begin{enumerate}
        \item   \textbf{E2E latency}: Based on the re-transmission probability and connection coverage, we design four types of E2E latency $\delta$ using the combination of these two factors (i.e., A and B for better and worse re-transmission probabilities respectively \cite{time-critical-comm}, 0.2 and 0.4 distance unit for the coverage in a 1 x 1 map). Generally, higher re-transmission probabilities induce longer delays, reducing the number of predictive trajectories sampled by edge servers. However, the sampling-based approach is more capable of handling the performance degradation as we mentioned in the introduction section. For the following two scenarios, the E2E latency setting $\delta$ is set to 1 as default (i.e., re-transmission probability A, 0.4 distance for the coverage).

        \item   \textbf{Density of edge servers}: Under the default E2E latency settings, we place different numbers of edge servers $N$ on the map to evaluate the performance in each run of this scenario. The random distribution of edge servers on the map follows a Voronoi diagram to delineate the connection coverage of edge servers by using polygonal boundaries, equally dividing the 1 x 1 map into multiple edge cloud regions. As the agent moves around and the surrounding areas change, the number of servers that the agent can connect to varies, resulting in different optimization levels. It is expected that with more servers available to assist, a more cost-optimal predicted trajectory can be calibrated collectively.

        \item   \textbf{Availability of edge servers}: In the above scenarios, we guarantee negligible average server loads $\mathcal{L}$ from the beginning to the end of the E-MPC operation so that servers are always available. Here we add one more uncertainty of the availability of edge servers for performance evaluation. Typically, as we analyzed in Section \ref{sec:analysis-multi}, the edge servers handle a bunch of requests from multiple agents simultaneously. For simplicity, in this scenario, the server availability follows a Bernoulli distribution with probabilities listed in Table \ref{table:parameter}. We leave the evaluation for a more complex scenario using the M/M/1 queuing model in the later section. As we described in Section \ref{sec:introduction}, the sampling-based MPC method is expected to have a graceful performance degradation when remote computational nodes are not available.
    \end{enumerate}
    
    \item   \textbf{Localized sensing information}: 
    In this scenario, we place the high-cost regions onto the reference path to evaluate performance in terms of unexpected disturbances. As previously stated in Sec. \ref{sec:analysis-local}, the agent pessimistically assumes the whole map is high-cost, while the edge node evaluates costs using accurate state information to help the agent avoid the high-cost region. We design two high-cost regions to reflect real-world scenarios due to unexpected environmental changes (e.g., snow, flood, rainfall, etc.). Furthermore, we conduct two motion-primitive libraries at higher and lower speeds to mimic the behavior of system model $P$ when the agent enters icy roads and mud areas respectively. To isolate the impact of local information, in this experiment, we assigned the same sample budget to both the server and the agent and guaranteed connectivity.
    
    \item   \textbf{Localized observation histories}:
    As aforementioned in Sec. \ref{sec:analysis-data}, the cost-optimal motion-primitive library becomes more powerful and reliable when more cost-optimal prior driving data is collected. For each unique map, we collect the prior driving data from the 3600 runs of the above experiment scenarios in total (i.e., 18 scenario options, each with 200 iterations). Regarding the mapping function $f_\text{mapping}$ and the corresponding definition of anchor point $A$, we disregard the steering angles and round up the x-y coordinates to the second floating point precision. As we apply nine different weight values $\beta$ to the uniform and the empirical distribution, we generate nine cost-optimal libraries using different amounts of collected data $\Omega$. The values of parameters $\beta$ and $\Omega$ are listed in Table \ref{table:parameter}. The evaluation results are plotted into a heat map with the color gradient indicating the performance (i.e., the total cost-to-go). It is expected that the lowest cost-to-go is located in using the most collected data but not completely relying on the empirical distribution formed by the cost-optimal prior data (i.e., $\beta$ equals to 1), as stated previously in Sec. \ref{sec:data-traj}. To isolate the impact of local observation histories in this evaluation, we assigned the same sample budget to both the server and the agent again and guaranteed connectivity. Note that same as the previous scenario, the agent only connects to one edge server.

    \item   \textbf{Multi-agent scenarios}:
    To evaluate the impact of server load against the performance of E-MPC in a multi-agent scenario, we first conduct an analytical simulation to simulate the delay time of each edge server (i.e., wait time plus service time) using the M/M/1 queuing model as described in Sec. \ref{sec:analysis-multi}. The server delay is then embedded into the E-MPC framework to further affect the number of trajectory samplings sent from edge servers to the agent. Following the same evaluation method in the previous scenario, we plot the evaluation results into heat maps with nine parameter pairs of M/M/1 queuing model ($\lambda$, $\mu$) on the x-axis and nine different numbers of edge servers on the y-axis. However, since the goal of this scenario is to evaluate how many edge servers need to be deployed to fulfill the service requirements (e.g., performance and reliability), we plot two different heatmaps. The color gradient of the first heat map represents the total cost-to-go, while the second one shows the probability of failure (PoF). As we sort the parameter pairs in an ascending order using the mean sojourn time in (\ref{eqn:sojourn}), it is expected that fewer edge servers are required to fulfill a certain level of service requirement.

    \item   \textbf{Random maps}:
    For better presentation, the above evaluations only consider 2 or 4 unique maps. Thus, to validate the generality of the E-MPC framework, we select 40 unique maps from 100 different random seeds. The evaluation scenarios include 1) the increased computational capacity and 2) the wireless communication using the same parameter settings.

    \begin{figure}[ht!]
    \centering
    \subfloat[]{%
        \includegraphics[width=0.5\linewidth]{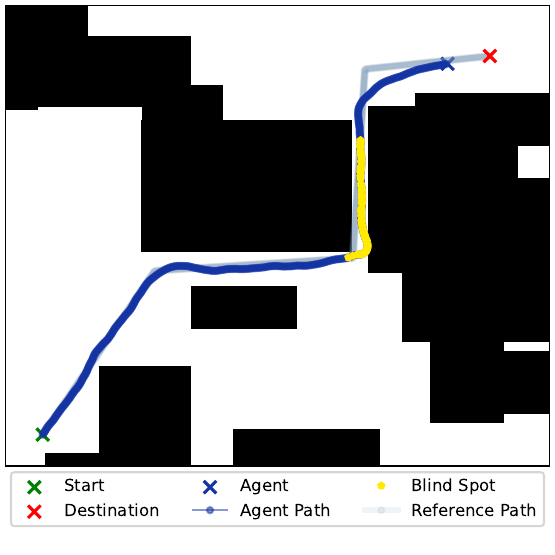}
        \label{4a}}
    \subfloat[]{%
        \includegraphics[width=0.5\linewidth]{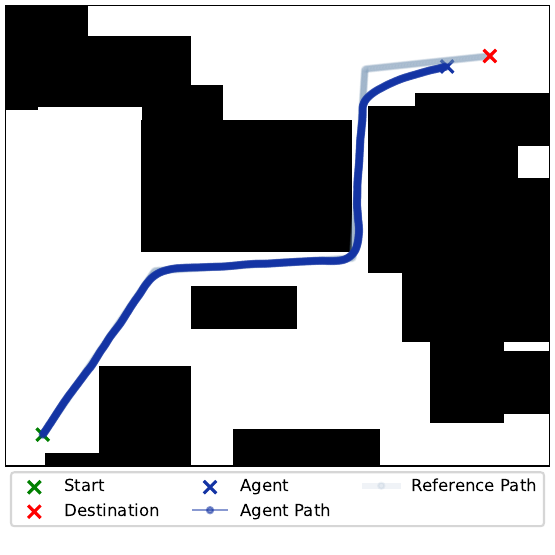}
        \label{4b}}
    \vfill
    \subfloat[]{%
        \includegraphics[width=0.5\linewidth]{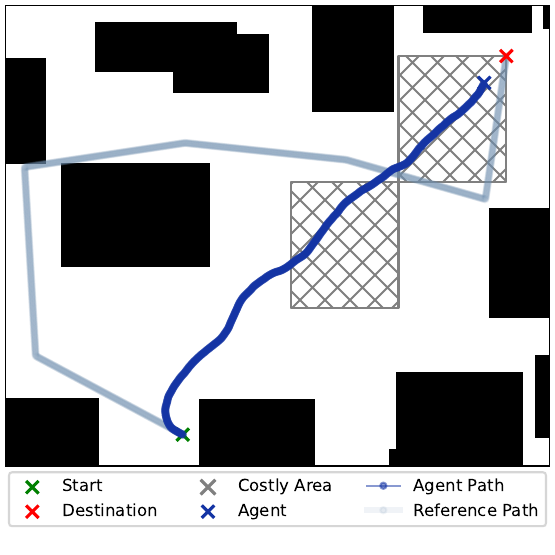}
        \label{4c}}
    \subfloat[]{%
        \includegraphics[width=0.5\linewidth]{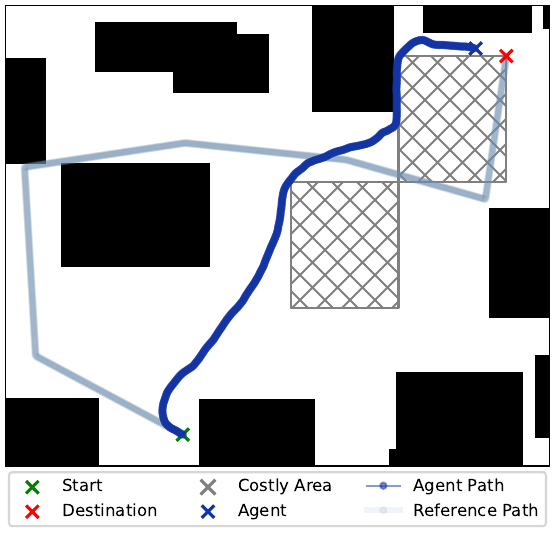}
        \label{4d}}
    \\
\caption{Comparing standard MPC and E-MPC from visualized map view which highlights the reduction in blind spots and high-cost region avoidance (a) Map 160: Standard MPC (b) Map 160: E-MPC (c) Map 878: Standard MPC (d) Map 878: E-MPC.}
\label{fig:maps}
\end{figure}
    
\end{enumerate}


\begin{figure*}[ht!]
\centering
    \centering
    \subfloat[]{%
        \includegraphics[width=0.33\linewidth]{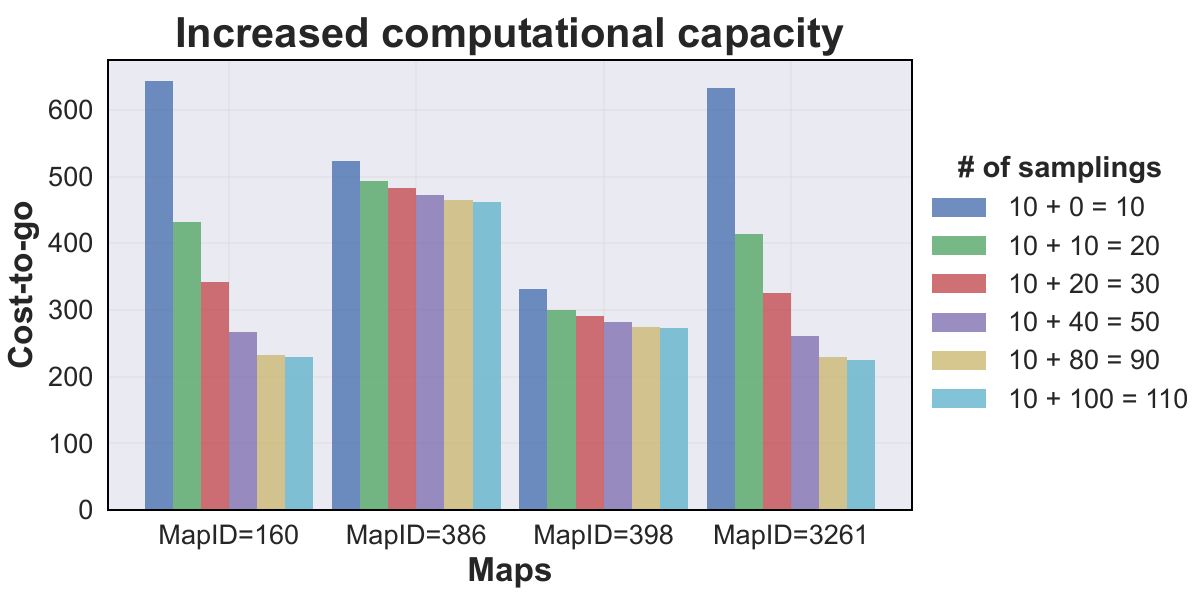}
        \label{5a}}
    \subfloat[]{%
        \includegraphics[width=0.33\linewidth]{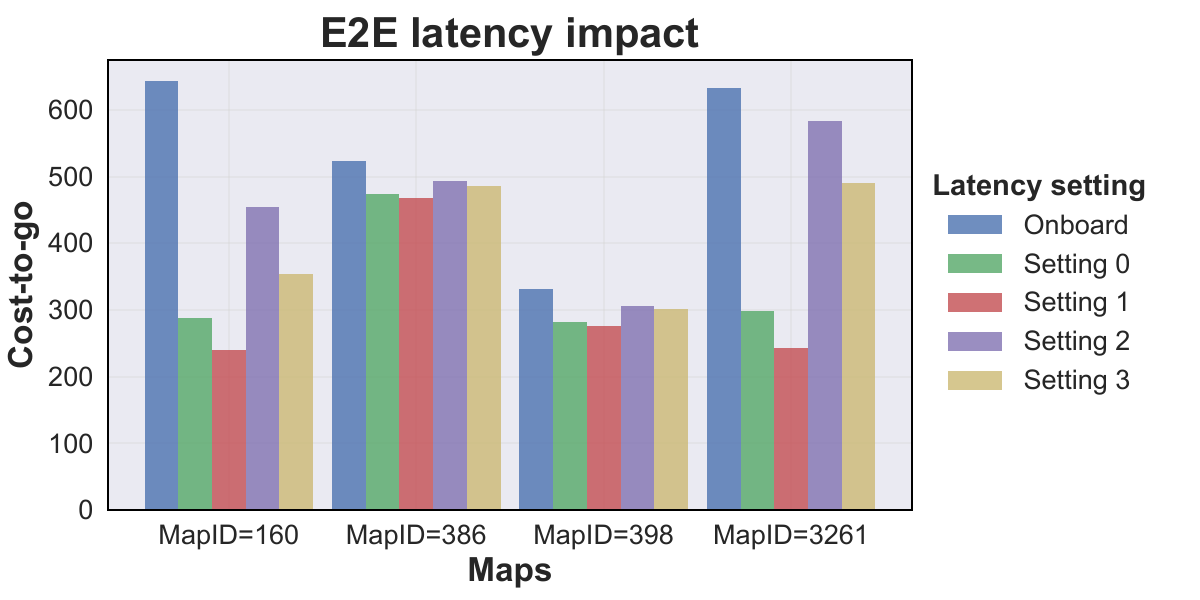}
        \label{5b}}
    \subfloat[]{%
        \includegraphics[width=0.33\linewidth]{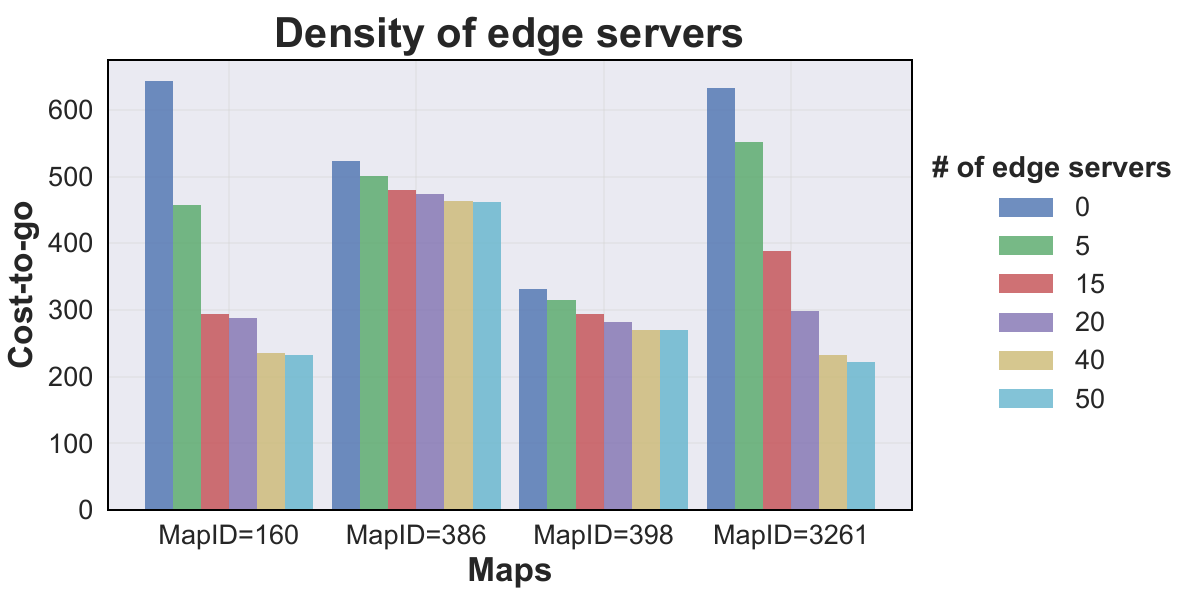}
        \label{5c}}
    \\
    \subfloat[]{%
        \includegraphics[width=0.33\linewidth]{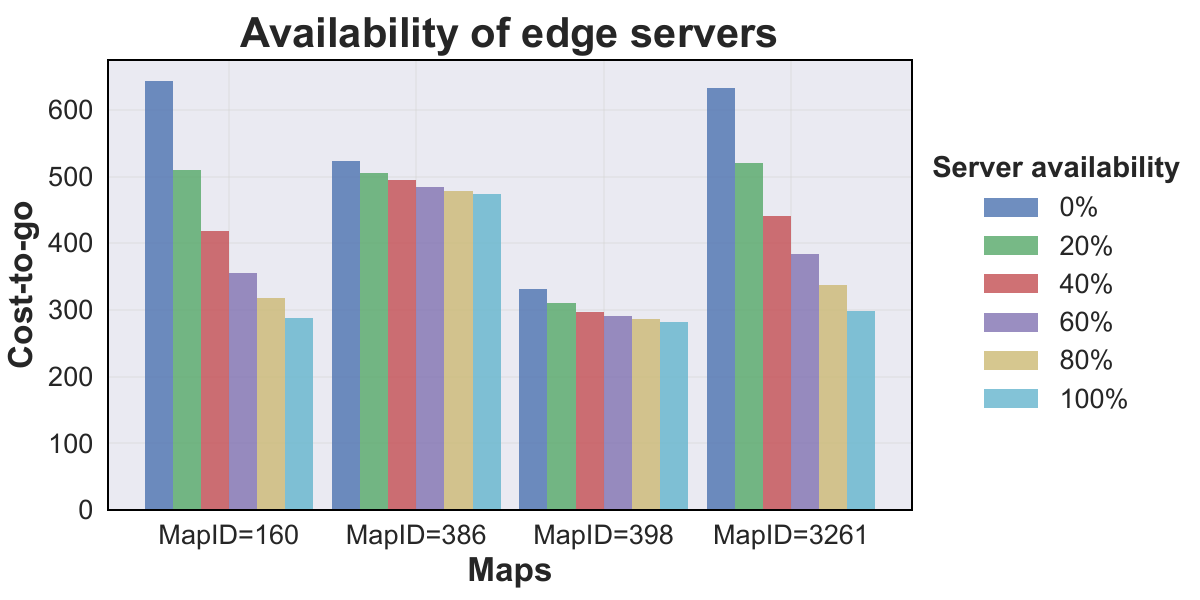}
        \label{5d}}
    \subfloat[]{%
        \includegraphics[width=0.33\linewidth]{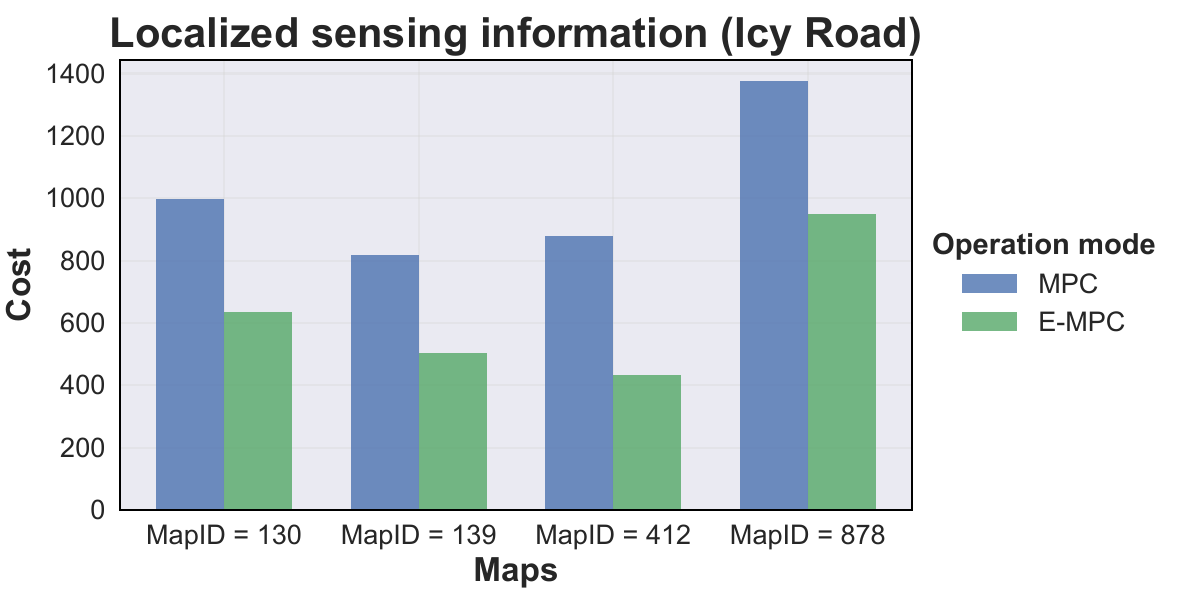}
        \label{5e}}
    \subfloat[]{%
        \includegraphics[width=0.33\linewidth]{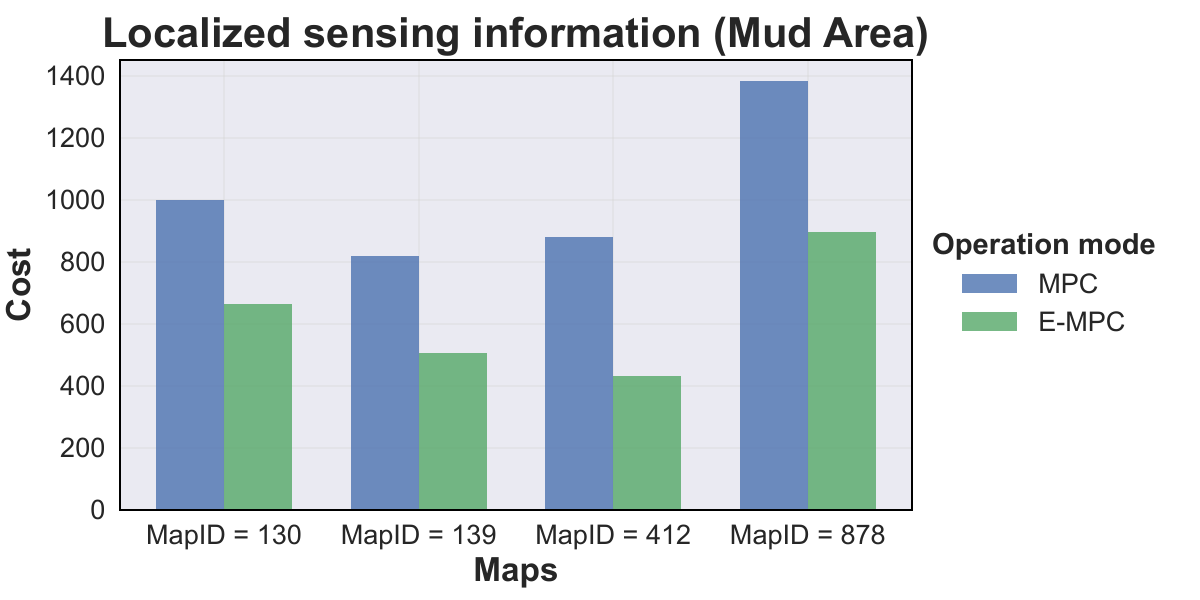}
        \label{5f}}
    \caption{Evaluation bar charts for the experiments: (a) Increased computational capacity (b) E2E latency impact (c) Density of edge servers (d) Availability of edge servers  (e) Localized sensing information: Icy road (f) Local sensing information: Mud area.}
    \label{fig:result-bar}
\end{figure*}

\subsection{Results} \label{sec:eval-result}

We conduct a series of numerical evaluations to underline the benefits and challenges of E-MPC. Figure \ref{fig:maps} visualizes the performance improvement of E-MPC in the map view. Due to space constraints, we display four randomly generated maps with a zoomed-in perspective of the optimal driving path side-by-side. Then, for each subsequent analysis, we present the corresponding numerical results in bar plots and heat maps along with the intuitive analysis based on the map view. We run MPC and E-MPC for 200 iterations for each option in the evaluation before averaging the results.

Note that we only consider the valid paths in the evaluation, where a valid path is defined as one in which the agent reaches a fixed and close-enough distance from the destination (i.e., 0.03 distance unit in a 1 x 1 map) without colliding with obstacles. When the agent is close enough to the destination, assistance from the edge server is no longer required.

\subsubsection{\textbf{Increased computational capacity}}


To assess the impact of increased computational capacity, we assign the agent a fixed computational capacity to perform standard MPC. We similarly set a fixed computational capacity for the edge server, but we increase it incrementally as stated in Sec \ref{sec:eval-scenario}. Figure \ref{4a} depicts a typical yet risky scenario in which an agent positioned on a narrow road frequently encounters a blind spot due to insufficient sampling, resulting in a high-cost route. In contrast, with more E-MPC sampling of the predicted trajectory in Fig. \ref{4b}, the agent discovers a cost-optimal motion primitive that prevents the agent from entering a blind spot. Numerical analysis is depicted in bar charts in Fig. \ref{5a}. The cost of a valid path is lowered by up to 65 percent when the edge server can conduct 100 E-MPC samplings within a fixed re-planning rate.

\subsubsection{\textbf{E2E latency in edge computing}}

We evaluate the effect of wireless communication dynamics by configuring an agent’s E2E latency settings differently for each run. Given that the E2E latency settings are dependent on the influence of obstacles and connection distance, 20 edge servers are randomly distributed on the map. The underlying map view is referred to as the system framework figure. Figure \ref{5b} shows that E-MPC outperforms standard MPC in terms of cost savings by up to 62.7\%. Although the cost-reduction level decreases in a higher re-transmission probability and shorter coverage (i.e., E2E latency setting 2), E-MPC still outperforms standard MPC by at least 8\%.

\begin{figure}[ht!]
    \centering
    \includegraphics[width=0.45\textwidth]{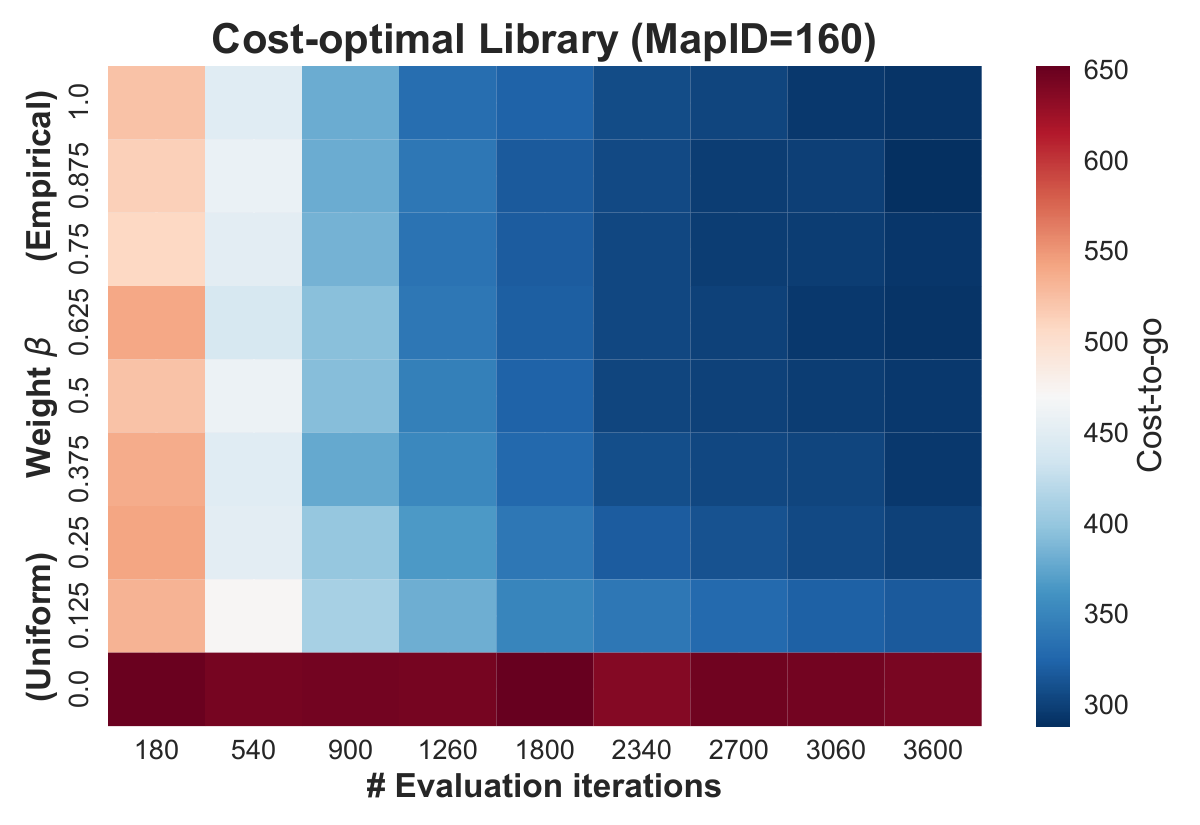}
    \caption{Heat maps for the experiment localized observation histories evaluated using nine weight parameter $\beta$ and nine cost-optimal prior libraries.}
    \label{fig:result-heatmap-1}
\end{figure}

\subsubsection{\textbf{Density of edge servers}}


As more edge servers are deployed to assist the agent, E-MPC demonstrates an improved route that has a more stable steering control navigating toward the destination and keeps the appropriate distance from the obstacles. The agent is also able to navigate through a right-angle lane without entering blind spots while respecting a sub-optimal reference path. The visualized improvement is similar and can be referred to in Fig. \ref{4a} and Fig. \ref{4b} due to the page limitation. The numerical evaluation results are plotted in Fig. \ref{5c}. The lowest cost-to-go is close to the best one in the evaluation of increased computational capacity with up to 63.7\% cost-reduction level. Note that in the case of map 160, the performance improvement is not strictly monotonic when more edge servers are deployed. In this case, the realistic impacts of wireless dynamics cause the agent to receive a similar assistance level from the edge when there are 15 and 20 servers.

\subsubsection{\textbf{Availability of edge servers}}

In addition to the varying computational capacity of edge servers, a fixed probability is assigned to the server availability that is modeled as Bernoulli distribution. The availability of edge servers is determined at the beginning of each move of an agent. To isolate the impact of the server availability, we randomly distributed 20 edge servers on each unique map. Figure \ref{5d} demonstrates that the performance of E-MPC falls gracefully as more edge servers become unavailable.

\subsubsection{\textbf{Localized sensing information}}


To determine the effect of undetectable disturbances on the agent, we insert two high-cost regions at two distinct positions along the reference path shown in Fig. \ref{4c} and \ref{4d}. Compared to Fig. \ref{4c}, Fig. \ref{4d} shows that with assistance from the edge servers, the agent is able to decide to avoid or enter the high-cost regions smartly. The numerical results are shown as bar charts in Fig. \ref{5e} and \ref{5f} for two environmental impacts (i.e., icy roads and mud areas). As agents enter and stay in high-cost regions longer in the standard MPC case, the total cost-to-go is much higher than E-MPC.

\begin{figure}[ht!]
\centering
    \centering
    \subfloat[]{%
        \includegraphics[width=0.5\textwidth]{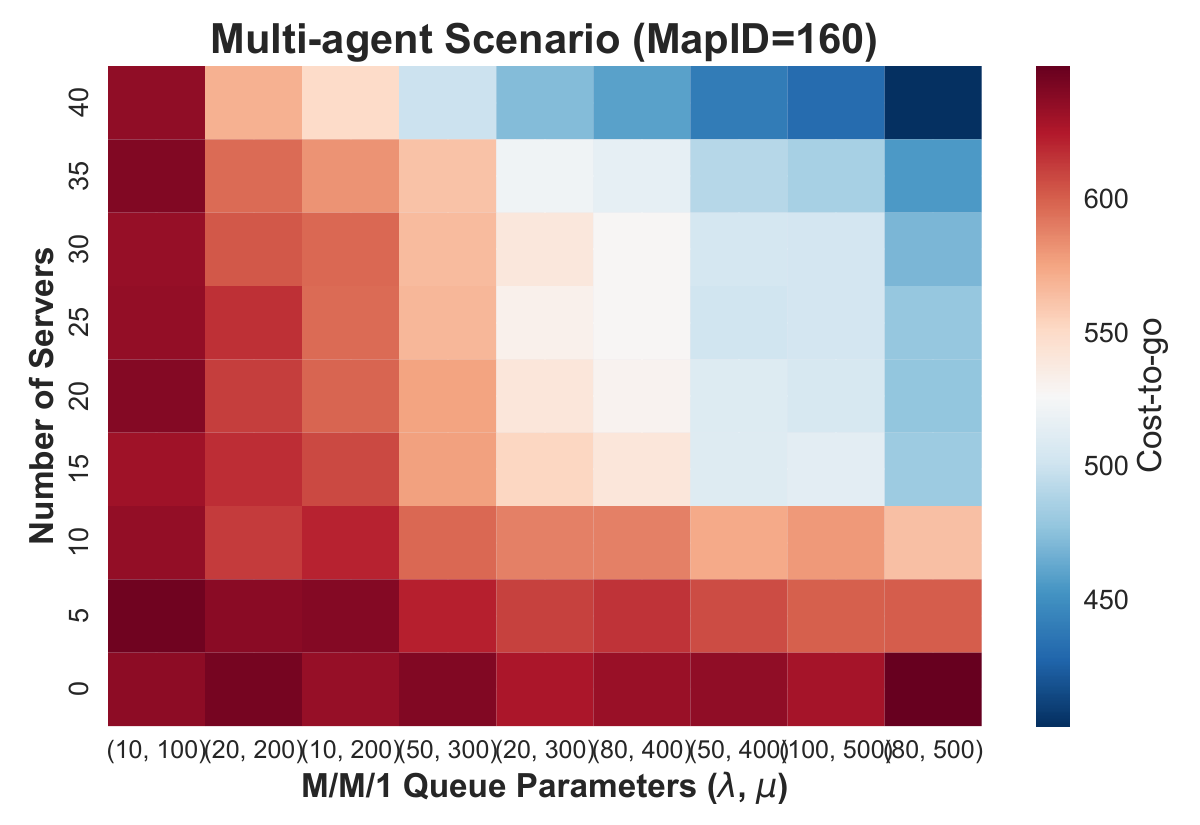}
        \label{7a}}
    \vfill
    \subfloat[]{%
        \includegraphics[width=0.5\textwidth]{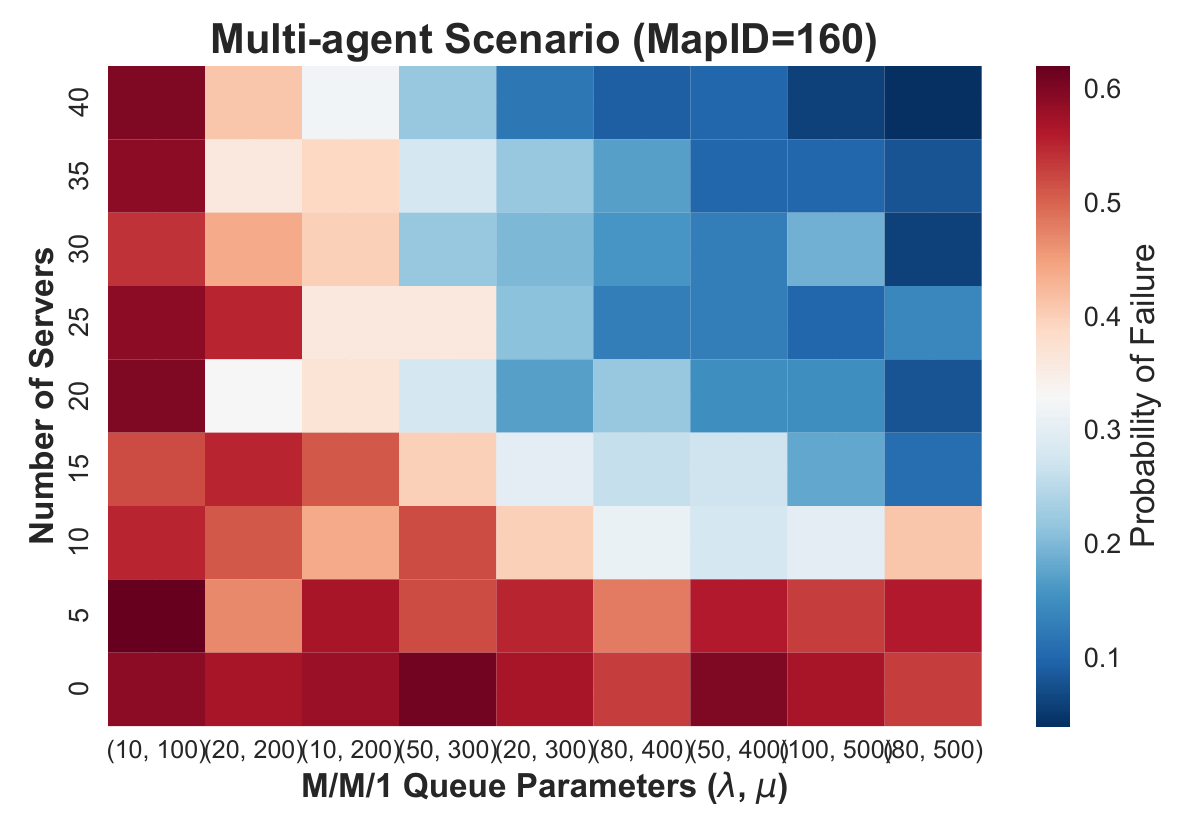}
        \label{7b}}
    \caption{Evaluation of multi-agent scenario against service requirements in (a) Map 160: Cost-to-go as performance, and (b) Map 160: Probability of failure (PoF) as reliability.}
    \label{fig:result-heatmap-2}
\end{figure}

\subsubsection{\textbf{Localized observation histories}}

Figure \ref{fig:result-heatmap-1} depicts the evaluation heat maps for two different maps. When the weight parameter $\beta$ is set to zero as in the last row, the sampling distribution equals the default uniform distribution, resulting in the worst performance. On the other hand, when the $\beta$ is set to one as in the first row, the sampling distribution completely follows the empirical distribution. However, we can find that the lowest cost-to-go is not in the first row, and by considering uniform distribution jointly, the agent is able to find the best route. This result validates our statement in Sec. \ref{sec:benefit-3} that it is risky to rule out potentially optimal actions even though the motion primitives have not been sampled yet. Lastly, we highlight the non-monotonic optimization trend regarding different $\beta$ values in a fixed amount of collected data. Since the continuously accumulating cost-optimal prior data forms different empirical distributions, one should keep searching for the best $\beta$ value. However, due to the large state space of real numbers, it can be computation-intensive and time-consuming to reach the best performance by dynamically and adaptively adjusting the weight parameter $\beta$.

\subsubsection{\textbf{Multi-agent scenarios}}

The evaluation results of the multi-agent modeling are plotted into heat maps as shown in Fig. \ref{fig:result-heatmap-2}. When the server load gets higher from the rightmost column to the leftmost one, more edge servers are needed to be deployed to maintain a service-level agreement (SLA). Take map 160 for instance, to guarantee a cost-to-go lower than 550 and a PoF lower than 0.4, we need to deploy at least 20 servers. However, when the average server load is too high to assist the agent, deploying 40 servers is still not enough to satisfy the SLA. Although the M/M/1 queuing model might be too simplistic in the real world, this numerical analysis provides a preliminary estimation for the trade-offs between performance and server deployment cost.

\subsubsection{\textbf{Random maps}}

\begin{figure*}[ht!]
\centering
    \centering
    \includegraphics[width=1\linewidth]{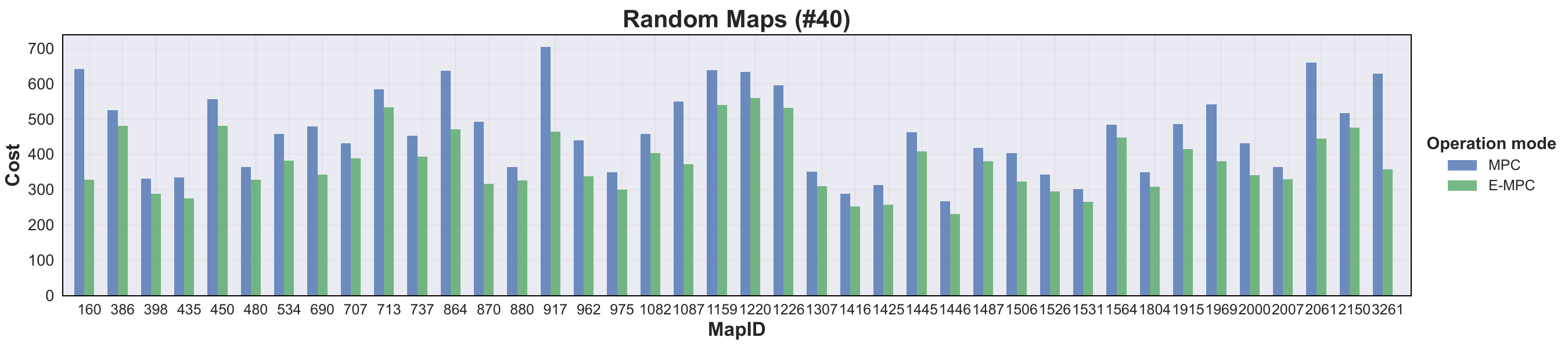}
    \caption{Evaluation chart for 40 random maps with the average cost-to-go of 18 scenario options as the performance indicator.}
    \label{fig:result-random-maps}
\end{figure*}

Figure \ref{fig:result-random-maps} shows the average cost-to-go of 40 random maps over the 18 scenario options as described in Sec. \ref{sec:eval-scenario}. Although the cost-reduction level is negligible around 10\% for some maps (e.g., 880, 2007), the overall improvement proves that the E-MPC framework employs realistic settings and thus can be applied to different environments generally, not meticulously contrived.

%% file: 7-Conclusion.tex
\section{Conclusion} \label{sec:conclusion}


We have proposed a promising method for edge-assisted MPC (E-MPC) as a structural improvement for learning-based control. The heterogeneity of the edge network provides clear benefits in terms of its low-latency computational assistance, localized sensor information, and localized observation histories. These capabilities have the potential to offer dramatic improvements to model predictive control, especially in the case of agents operating in an environment for the first time. Moreover, we evaluate the needs of computational resources for achieving service requirements using a multi-agent model to estimate the average server loads. Autonomous vehicle (AV) operators can benefit from such analysis in calculating computing and server deployment costs by reflecting the actual loads of their cloud servers in production. Future works include but are not limited to considering more complicated interactions between multiple AVs on the map. When the reference paths overlap and the AVs encounter each other, the global and local planners should cooperate to find the best detour route without collision. 
In addition, it is possible to implement the E-MPC framework and conduct real experiments using the MuSHR/ROS car platform. 
Lastly, the trade-offs associated with each benefit are also crucial to investigate further, such as hardware costs and management against increased computation and heterogeneous edge servers. Also, data security and storage limitations are important for storing localized sensing information and historical data.

%% file: 8-Proof.tex
{\appendix[A. Utilizing Priors in Sampling-based Cost Minimization]\label{sec:appendix}

Considering that the several points $D_{\text{points}}$ can be sampled at anchor points $A$ from the motion-primitive library in the space $X$, the spatial relation can be written as $D_{\text{points}} \in D_{\alpha} \subset X_{\text{space}}$. The average performance (cost-to-go) can be calculated:
$$\mu = \frac{1}{\lvert D_{\alpha} \lvert} \sum_{\alpha^{(i)} \in D_{\alpha}} C(\alpha^{(i)} \lvert x).$$
For $M$ driving iterations with $N \ll \lvert D_{\alpha} \lvert$ samplings per moving step: (1) For the i-th iteration that the agent enters an anchor point $x$ without historical data, the probability of getting a trajectory with a cost-to-go lower than the mean value $\mu$ can be described as $p_x (C(\alpha^{(i)} \lvert x) \leq \mu) = P_{\mu^{-}} = \frac{\lvert D_{-} \lvert}{\lvert D_{\alpha} \lvert},$ where $D_{-}$ is a set of points on the trajectories with better performance compared to the mean value $\mu$. Thus, the lowest possibility to outperform average cost-to-go with consecutive $N$ sampling is defined as $\delta_{1} = 1 - (1 - P_{\mu^{-}})^N.$ (2) For the m-th iteration that the agent enters an anchor point $x$ stored with several historical data, the probability of getting a better trajectory is expected to be higher than $\delta_{1}.$ Assuming several points $m_{-}$ with better performance are collected, the trajectory sampling now belongs to the cost-optimal library, $\alpha^{(j)} \in D'_{\textit{prior}}.$ Then, the probability of getting a trajectory with a cost-to-go lower than the mean value can be described as $p_x (C(\alpha^{(j)} \lvert x) \leq \mu) = P'_{\mu^{-}} =\frac{m_{{-}}}{m}.$ Thus, the lowest possibility to outperform average cost-to-go with consecutive $N$ sampling is defined as $\delta_{2} = 1 - (1 - P'_{\mu^{-}})^N.$ If $\delta_{2}$ is higher than $\delta_1$, then the amount of priors that need to be collected can be estimated as $\delta_{2} = 1 - (1 - P'_{\mu^{-}})^N \geq \delta_{1} = 1 - (1 - P_{\mu^{-}})^N, \frac{m_{{-}}}{m} \geq \frac{\lvert D_{-} \lvert}{\lvert D_{\alpha} \lvert}.$

}